\documentclass{vgtc}                          

\graphicspath{{figures/}{pictures/}{images/}{./}} 

\usepackage{times}                     

\usepackage{tabu}                      
\usepackage{booktabs}                  
\usepackage{lipsum}                    
\usepackage{mwe}                       
\usepackage{enumitem}
\usepackage{mathptmx}                  

\onlineid{0}

\vgtccategory{Research}

\vgtcinsertpkg
\usepackage{orcidlink}

\title{Exploring the Feasibility of Gaze-Based Navigation Across Path Types}

\author{
Yichuan Zhang\textsuperscript{1,2}~\orcidlink{0009-0008-1142-1501}\thanks{Both contributed equally to this work.}
\and
Liangyuting Zhang\textsuperscript{2}~\orcidlink{0009-0007-6255-6513}\footnotemark[1]
\and
Xuning Hu\textsuperscript{1,2}~\orcidlink{0009-0009-1305-2081}
\and
Yong Yue\textsuperscript{2}~\orcidlink{0000-0001-7695-4538}
\and
Hai-Ning Liang\textsuperscript{1}~\orcidlink{0000-0003-3600-8955}\thanks{Corresponding author (e-mail: hainingliang@hkust-gz.edu.cn)}
}

\affiliation{\scriptsize \textsuperscript{1} Computational Media and Arts Thrust, The Hong Kong University of Science and Technology (Guangzhou), Guangzhou, China\\ 
\textsuperscript{2} School of Advanced Technology, Xi'an Jiaotong-Liverpool University, Suzhou, China\\
}

\abstract{
Gaze input, as a modality inherently conveying user intent, offers intuitive and immersive experiences in extended reality (XR). With eye-tracking now being a standard feature in modern XR headsets, gaze has been extensively applied to tasks such as selection, text entry, and object manipulation. However, gaze-based navigation—despite being a fundamental interaction task—remains largely underexplored. In particular, little is known about which path types are well-suited for gaze navigation and under what conditions it performs effectively. To bridge this gap, we conducted a controlled user study evaluating gaze-based navigation across three representative path types: linear, narrowing, and circular. Our findings reveal distinct performance characteristics and parameter ranges for each path type, offering design insights and practical guidelines for future gaze-driven navigation systems in XR.
} 


\begin{document}


\firstsection{Introduction}

\maketitle

In today's VR systems, users are typically offered multiple interaction paradigms, including bare-hand input and physical controllers, to support tasks such as selection~\cite{Xuning2025Selection, Difeng2019TOGSelection}, crossing~\cite{huang2020modeling, HuaweiCHI2021Crossing}, and navigation~\cite{zavichi2025gaze}. However, while gaze has become widely adopted for selection tasks—including Gaze-Pinch~\cite{pfeuffer2017gaze+}, multi-object selection~\cite{Kim2025CHIGazeSemiPinch}, and precision selection in cluttered environments~\cite{wei2023predicting}, its application in crossing and navigation tasks remains underexplored. Navigation, which involves guiding virtual objects along predefined trajectories from an origin to a goal, is central to applications such as digital gaming, hierarchical menu traversal, and driving simulations. To date, only Hu et al.~\cite{xuning2025CHILBW, HuTVCG2025GazeSteering} have investigated gaze-based steering in constrained straight-line paths. Yet real-world navigation scenarios are often more complex, involving curved or non-linear trajectories. Therefore, investigating diverse path parameters is essential for assessing the feasibility of gaze-based navigation and for developing robust, user-friendly gaze interaction techniques in immersive environments.

Trajectory-based navigation tasks—such as goal crossing and path steering—have long been foundational topics in human–computer interaction (HCI). Prior studies have systematically examined how trajectory parameters (e.g., curvature and width) affect user performance in graphical steering tasks~\cite{Yamanaka2016Narrowing}. Despite extensive research into both trajectory-based navigation and gaze interaction individually, their integration has received limited attention. Gaze input exhibits unique characteristics—such as inherent instability and different feedback dynamics—that challenge the direct transfer of findings from hand-based paradigms to gaze-driven contexts.

To address this gap, we conducted a controlled user study ($N=8$) to systematically evaluate the effects of path geometry on user performance in gaze-based navigation tasks. Participants completed navigation trials across three representative path types: Constant-width Linear Path, Narrowing Path, and Constant-width Circular Path. By analyzing objective metrics (e.g., movement time and re-entry count) alongside subjective workload ratings, this study aims to reveal the behavioral dynamics of continuous gaze-based navigation and offer actionable insights for the design of next-generation VR interfaces.

\section{Related Work}

With the increasing adoption of eye-tracking technologies, commercial Head-Mounted Devices (HMDs) such as the Apple Vision Pro, Meta Quest Pro, and HTC Vive Focus 3 have begun to integrate gaze tracking as a key input modality. As a maturing technology, eye tracking has been applied across a wide range of domains, including gaming~\cite{pakov2019collaborative, isokoski2009gazeGame}, text entry~\cite{Lui2021iText, Kurauchi2016GazeTextEntry, HedeshyCHI21GazeHumTextEntry}, target selection~\cite{meng2022textselection}, user interface design~\cite{Stellmach2012gazeui,piotrowski2019gaze}, and navigation~\cite{kang2024rayhand, Yushi25TVCGSteeringLatency, Yushi24TVCGSteeringDirection}---either as a standalone input method or as an auxiliary modality combined with other interactions.

Gaze-based interaction offers an intuitive, hands-free alternative to conventional input methods~\cite{Sidenmark2023Comparing, Majaranta2014Tracking, SibertCHI00GazeSelection}. This makes it especially attractive in scenarios where manual input is limited or undesired~\cite{Xueshi2021UISTHandsfree, Xueshi2020IsmarHandsfree}. However, it differs fundamentally from traditional pointer-based control. Scott et al.~\cite{Scott2004motor} attributed the higher re-entry rates in gaze-based input to intrinsic limitations of the gaze system---unlike regular cursor-based input, users cannot rely on continuous visual feedback for correction and must instead make predictive adjustments. Holmqvist et al. further observed that during correction phases, users tend to alternate between short saccades and fixations~\cite{Kenneth2011eyemovement}, ranging from 100 milliseconds to several seconds, stabilizing the cursor and using saccades to compensate for positional discrepancies~\cite{Prablanc1978saccades}.

While prior work has modeled gaze-based selection and steering tasks using adapted versions of Fitts’ Law and the Steering Law~\cite{xuning2025CHILBW, Zhang2010dwell-based}, research on gaze-based performance in more complex trajectory-based navigation tasks remains limited. Key path parameters, such as curvature and path narrowing, may significantly affect user performance in trajectory-based navigation~\cite{accot1997beyond}. Yet their influence on the gaze-based navigation task has not been systematically studied. Given that trajectory-based tasks differ notably from linear steering tasks in both motor behavior and cognitive demands, further investigation is needed to understand how these path parameters affect gaze-based navigation in immersive VR environments.

\begin{figure*}[htb]
  \centering
  \includegraphics[width=\linewidth]{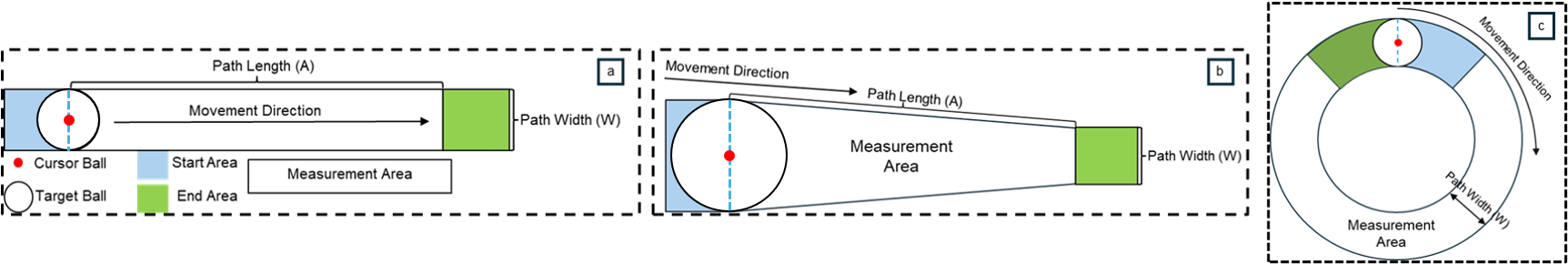}
  \caption{The red sphere indicates the user’s gaze cursor, while the green and blue spheres denote the task’s start and end positions, respectively. The path width is defined by the diameter of the target sphere. Participants manipulate the gaze cursor to guide the target sphere from the start area to the target area along the specified trajectory. (a) Constant-width Linear Path Task; (b) Narrowing Path Task; (c) Constant-Width Circular Path Task.}
  \label{F: Experiment Design}
\end{figure*}

\section{Experiment}
This study systematically investigates how different path geometries—specifically linear, narrowing, and circular trajectories—affect user performance and subjective experience during gaze-based navigation tasks. By evaluating these distinct path types under controlled conditions, we aim to uncover the suitability and limitations of gaze input for continuous motion control in immersive environments.

\subsection{Participants and Apparatus}
Eight participants (4 males and 4 females) were recruited from a local university. Their ages ranged from 19 to 52 years ($M = 32.75, SD = 14.08$). Using a 7-point Likert scale, participants rated their familiarity with VR systems ($M = 2.8, SD = 2.08$) and their familiarity with gaze-based interaction ($M =3.75, SD = 1.85$), with higher scores indicating greater familiarity. 

The experiment utilized the Meta Quest Pro headset, which features integrated eye-tracking capabilities as part of its standard hardware. Visual output was delivered at a resolution of 1800 × 1920 pixels per eye, with a horizontal field of view of approximately 106° and a vertical field of view of 95.57°. Gaze position data were sampled at a frequency of up to 72 Hz using Meta's official Unity Eye Tracking API \cite{Abeysinghe2025QuestProRefreshRate}. The system ran on a high-end desktop equipped with an Intel i9 processor and an NVIDIA RTX 3080 Ti GPU, ensuring smooth rendering and real-time data capture throughout the experimental sessions.

\subsection{Experimental Task}
To investigate user performance in different path-following scenarios, we designed three distinct task conditions~\cref{F: Experiment Design}: (a) Constant-width Linear Path, (b) Narrowing Path, and (c) Constant-Width Circular Path. At the start of each trial, participants pressed a button on the controller to activate eye tracking, enabling control of the red gaze cursor. The trial began once the cursor remained within the green starting zone for 500 ms, at which point the zone changed color to indicate initiation. Participants then guided a target ball—its diameter corresponding to the path width—along the prescribed path from the green start region to the blue end region. A trial was recorded as a re-entry if the gaze cursor moved outside the boundaries of the path, subsequently re-entered the boundaries, contacted the ball, and regained control. Participants were instructed to maintain a balance between speed and accuracy throughout the experiment.

\subsection{Design and Procedure}
In this experiment, we followed a within-subjects experiment design with three distinct path types: Constant-width linear path, Constant-width circular path, and Narrowing path. Each path type was parameterized as follows:

\textbf{Constant-width Linear Path:} In this condition, participants were required to steer the target sphere along a straight path of fixed width, see~\cref{F: Experiment Design}.a for a schematic illustration. The task was parameterized with two path lengths, $A$ ($30^\circ, 40^\circ$), and three path widths, $W$ ($3^\circ, 4^\circ, 5^\circ$). 

\textbf{Narrowing Path:} According to \cref{F: Experiment Design}.b, participants steered the target sphere along a path whose width gradually decreased from the start to the end area. The path is parametrized with length $A$ ($30^\circ, 40^\circ$), $W_\mathrm{Start}$($4^\circ, 5^\circ, 6^\circ$) and ending width $W_\mathrm{end}$ ($3^\circ$).

\textbf{Constant-width Circular Path:} Participants followed a circular trajectory, maintaining the target sphere within a path of constant width as they moved from the start to the end area, as shown in~\cref{F: Experiment Design}.c. Two levels of circumference ($C = 60^\circ$, $80^\circ$) and two path widths ($W = 4^\circ$, $5^\circ$) were used.

\subsection{Evaluation Metrics}

For each trial, we systematically recorded two key metrics to assess user performance and subjective experience: movement time and re-entry count.

\begin{itemize}[noitemsep, topsep=3pt, parsep=3pt, partopsep=3pt]
\item \textbf{Movement Time ($MT$):} The time elapsed from when the participant initiated the task in the start area to when the target ball reached the designated end region. $MT$ is a standard measure in HCI research for quantifying efficiency and characterizing behavioral patterns in steering tasks~\cite{Xuning2024Spatial}.

\item \textbf{Re-entry Count:} The number of times participants had to re-establish control of the cursor and target ball after losing contact. This metric reflects saccadic stability and the degree of control required under different conditions.

\end{itemize}

\subsection{Results}

\begin{figure}[htb]
  \centering
  \includegraphics[width=\columnwidth]{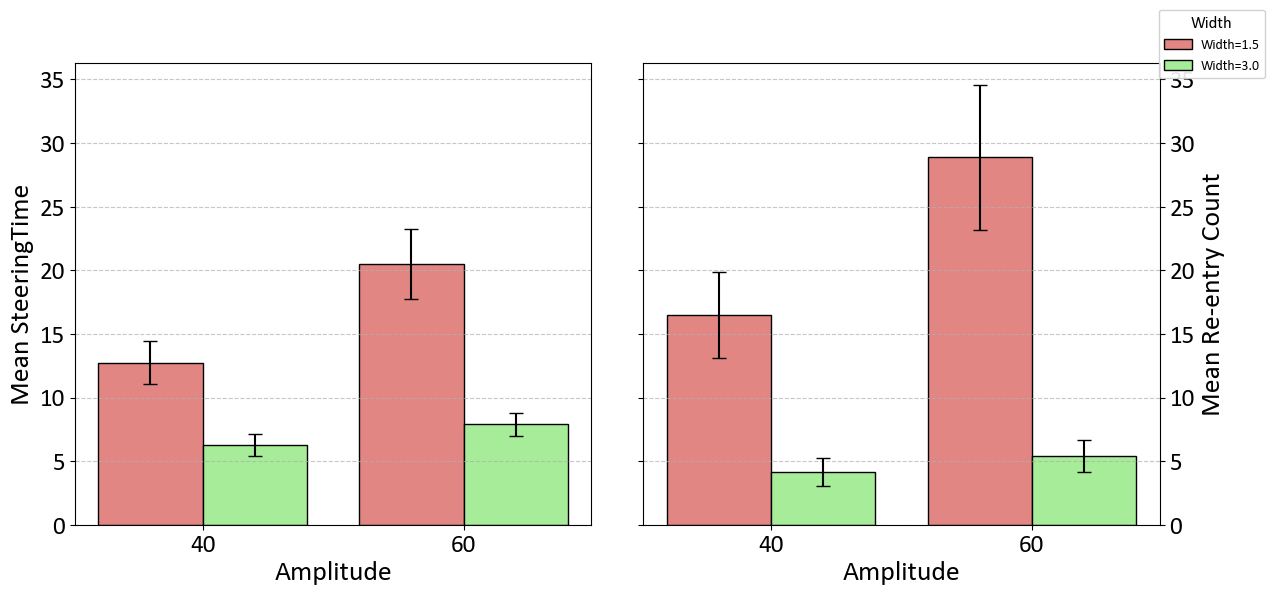}
  \caption{Circular path: mean steering time (left) and re-entry times (right) across amplitude and width. Error bars = SEM.}
  \label{fig:result-arc}
\end{figure}

\begin{figure}[htb]
  \centering
  \includegraphics[width=\columnwidth]{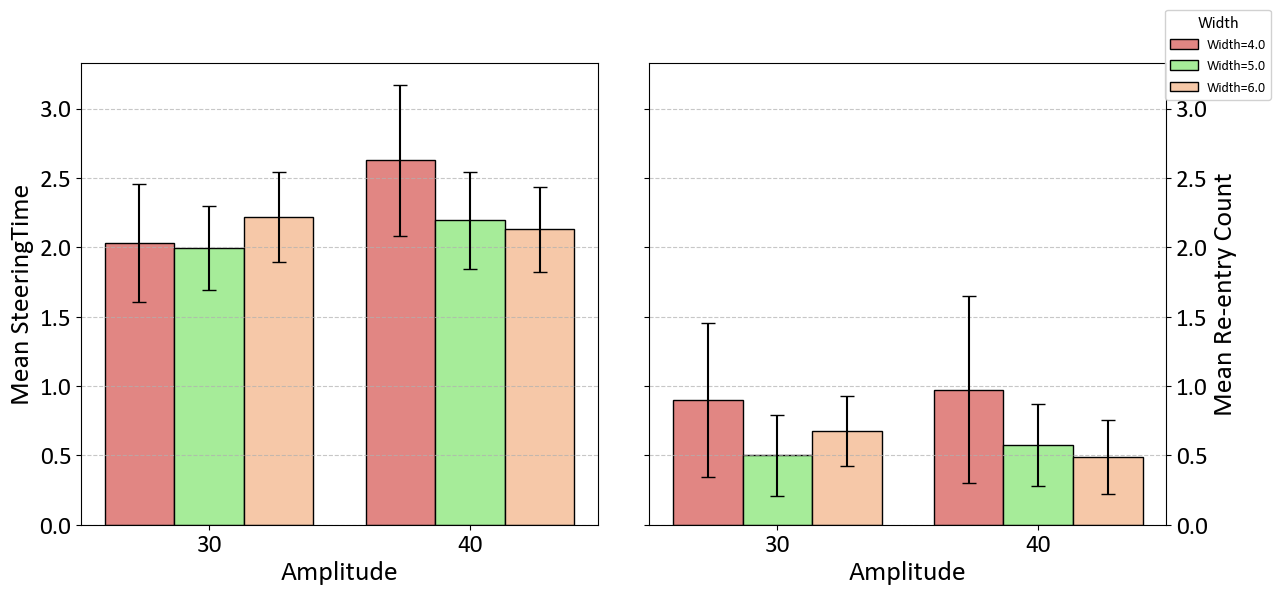}
  \caption{Narrow path: mean steering time (left) and re-entry times (right) across start and end widths. Error bars = SEM.}
  \label{fig:result-narrow}
\end{figure}

\begin{figure}[htb]
  \centering
  \includegraphics[width=\columnwidth]{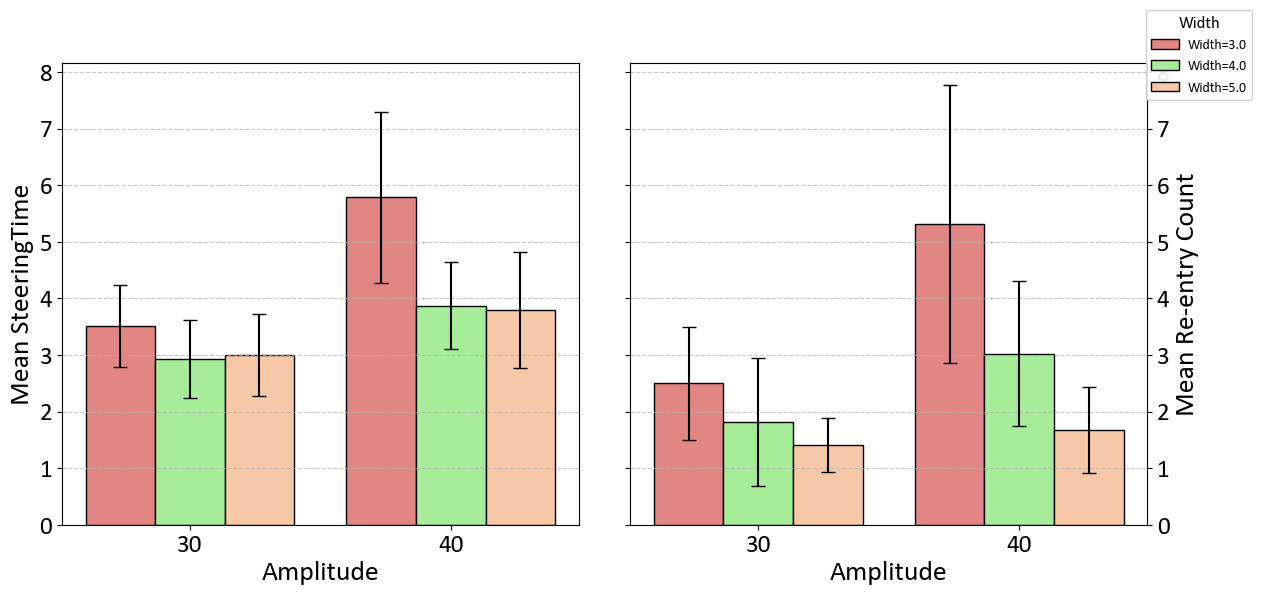}
  \caption{Rectangle path: mean steering time (left) and re-entry times (right) across circumference and width. Error bars = SEM.}
  \label{fig:result-circular}
\end{figure}

Repeated--measures ANOVA revealed, in the \textbf{Constant-width Linear Path} condition, no effects on Steering Time or Re-entry Times reached significance (all $p > 0.05$), although Amplitude showed a non-significant trend for Steering Time ($F(1,7) = 4.15$, $p = 0.081$, $\eta_p^2 = 0.06$). 

In the \textbf{Narrowing Path} condition, only the \textit{Amplitude}\,$\times$\,\textit{Width} interaction on Steering Time reached significance ($F(2,14) = 4.79$, $p = 0.026$, $\eta_p^2 = 0.02$); neither main effect was significant (all $p > 0.05$), and no effects were significant for Re-entry Times (all $p > 0.05$).

In the \textbf{Constant-Width Circular Path} condition, significant main effects of \textit{Amplitude} on \textit{Steering Time} ($F(1,7) = 31.29$, $p = 0.0008$, $\eta_p^2 = 0.23$) and \textit{Width} ($F(1,7) = 36.84$, $p = 0.0005$, $\eta_p^2 = 0.56$), as well as a significant \textit{Amplitude}\,$\times$\,\textit{Width} interaction ($F(1,7) = 17.91$, $p = 0.0039$, $\eta_p^2 = 0.11$). For \textit{Re-entry Time} in Constant-Width Circular Path, there were also significant main effects of Amplitude ($F(1,7) = 16.23$, $p = 0.0050$, $\eta_p^2 = 0.12$) and Width ($F(1,7) = 23.67$, $p = 0.0018$, $\eta_p^2 = 0.50$), and a significant interaction ($F(1,7) = 15.93$, $p = 0.0053$, $\eta_p^2 = 0.08$).

Overall, these results indicate that the Constant-Width Circular Path yielded strong amplitude and width effects (and their interaction) on both speed and accuracy, the Narrow path produced only an interaction effect on speed, and the Rect path showed weak or non-significant effects.

\subsection{Discussion}
Our study provides the first systematic comparison of gaze‑based navigation across three canonical path geometries—linear, narrowing, and circular—within a unified experimental framework. The results reveal that path curvature and width jointly modulate both efficiency and control demands. Specifically, the \textbf{Constant-Width Circular Path} condition exhibited pronounced main effects and an interaction between amplitude and width on movement time and re-entry times, indicating that gaze steering along curved trajectories is highly sensitive to both path length and available tolerance. This contrasts with the \textbf{Narrowing Path} condition, where only an amplitude–width interaction affected speed, and the \textbf{Constant-Width Linear Path} condition, where neither factor reached significance. Taken together, these findings suggest that curvature amplifies the difficulty imposed by constrained widths, whereas simple linear steering remains comparatively robust to moderate geometric variations.

Designers of VR applications should treat path curvature as a primary parameter when mapping gaze input to continuous motion. Wider lanes (i.e., $\geq 5^\circ$) substantially mitigated performance costs on curved trajectories, implying that generous visual margins are essential whenever users must negotiate turns using gaze alone. Conversely, merely widening the start of a narrowing path offered limited benefit, underscoring the importance of maintaining sufficient clearance along the entire route rather than at endpoints only. These insights extend prior work on gaze-based straight-line steering~\cite{xuning2025CHILBW} by showing that curvature, not narrowing per se, is the dominant source of instability under continuous gaze control.

\section{Limitations and Future Work}
This study highlights the limitations of using gaze as an input modality for navigation tasks and outlines several directions for future research. First, our experimental design focused on a set of representative path types—linear, narrowing, and circular—that reflect common patterns in XR environments. However, real-world navigation paths are often more complex and dynamic, suggesting the need to extend future investigations to irregular, branched, or adaptive trajectories.

Second, a notable limitation observed in our study was the frequent occurrence of re-entry events caused by the instability of gaze input. In practical scenarios, such re-entries may be interpreted as errors, compromising task performance and user experience. Addressing this issue requires further exploration into more robust eye-tracking technologies or the development of algorithmic techniques to enhance gaze stability, particularly in foundational tasks such as navigation.

\section*{Acknowledgment}
The authors thank the participants for their time and the reviewers for their comments, which helped improve our paper. This work was supported in part by the Suzhou Municipal Key Laboratory for Intelligent Virtual Engineering (\#SZS2022004).

\bibliographystyle{abbrv-doi}

\bibliography{template}
\end{document}